\documentclass[%
 reprint,
 amsmath,amssymb,
 aps,
floatfix,
longbibliography
]{revtex4-2}

\usepackage{xcolor}
\usepackage{graphicx}
\usepackage{float}
\usepackage{hyperref}

\newcommand{\Eq}[1]{Eq.~\ref{#1}}

\newcommand{\Fig}[1]{Fig.~\ref{#1}}

\newcommand{\partFig}[2]{Fig.~\hyperref[#1]{\ref*{#1}#2}}
\newcommand{\partFigure}[2]{Figure~\hyperref[#1]{\ref*{#1}#2}}

\begin{document}

\preprint{prx}

\title{
Irradiation-driven Evaporation of Micro Droplets in an Optical Trap
}

\author{Jugal Rakesh Shah}
\affiliation{Department of Mechanical Engineering, Indian Institute of Technology Indore, India\\}

\author{Max Huisman}
\affiliation{SUPA and School of Physics and Astronomy, The University of Edinburgh, Peter Guthrie Tait Road, Edinburgh EH9 3FD, United Kingdom\\}

\author{Devendra Deshmukh}
\affiliation{Department of Mechanical Engineering, Indian Institute of Technology Indore, India\\}

\author{Dag Hanstorp}
\affiliation{Department of Physics, University of Gothenburg, SE-412 96 Gothenburg, Sweden\\}

\author{Javier Tello Marmolejo}
\email{javier.marmolejo@physics.gu.se}
\affiliation{Department of Physics, University of Gothenburg, SE-412 96 Gothenburg, Sweden\\}

\date{\today}

\begin{abstract}

Small droplets are irradiated with visible and infrared light in many natural and industrial environments. 
One of the simplest ways to describe their evaporation is the D$^2$-Law.
It states that the evaporation rate is proportional to $t^{-1/2}$, and $R^{-1}$.
However, models like the D$^2$-Law do not account for the volumetric heating of light and the effect of strong irradiation on individual droplets is not fully understood.
Here we show the effects of IR irradiation on optically levitated water droplets.
We find that, under strong irradiation of up to $10^8 W/m^2$, the droplet evaporation is initially driven by the heat from the laser following the power law $dR / dt \sim R$, i.e. the inverse of the D$^2$-Law.
Then, when the droplets shrink to 2 - 3 $\mu$m in radius a turnover occurs from irradiation-driven back to diffusion-driven evaporation.
Our findings support the understanding of droplet evaporation in cases such as rocket engines or internal combustion, where the radiation from the flame will heat water and fuel droplets.

\end{abstract}

\maketitle

\section{Introduction}

The evaporation of small liquid droplets into the surrounding gas phase is ubiquitous in many industrial and scientific applications. 
Spray drying is often employed to produce fine particles for medical use \cite{Vehring2008,dugas_droplet_2005}, while in combustion science it is important to understand how fuel droplets evaporate in order to improve engine efficiency \cite{ray_evaluation_2023,kotake_evaporation_1969}. 
Other applications include atmospheric sciences \cite{sezen_water_2023} and biomedical research \cite{SEFIANE2010S82}. 
In some scenarios, droplets are subjected to radiative heating, which significantly influences the evaporation dynamics. 
These include fuel droplets inside a combustion chamber or rocket engine, where droplets are exposed to flames \cite{WANG2023108306}, cloud formation by aerosol droplets \cite{leung_aerosolcloud_2023}, where droplets are exposed to radiation from the sun, and laser diagnostics of fuel droplets \cite{ray_evaluation_2023} in combustion, where atomized fuel droplets are studied using optical diagnostics like LIF (Laser-Induced Fluorescence). 

To study the evaporation of individual droplets, many studies have relied on methods such as free-falling droplets \cite{tripathi_evaporating_2015}, sessile droplets on flat surfaces \cite{Jiao_2016}, and fiber-based studies \cite{duprat_evaporation_2013}. 
However, falling droplets induce convective flows \cite{barmi_convective_2014} due to the relative motion between the droplet and the ambient medium, and the solid surfaces introduce heat conduction and imperfect sphericity of the droplets. 

To mitigate these limitations, various non-contact levitation methods have been developed for evaporation studies, such as electrostatic \cite{jin_electrostatic_2022} and acoustic traps \cite{ali_al_zaitone_evaporation_2011}. 
However, these methods also face significant challenges. 
Electrostatic traps, for instance, require droplets to be charged. 
This introduces issues of charge saturation \cite{jakubczyk_investigation_2004}, which can cause droplets to break up \cite{Li_2005} and sets a lower limit to their size to approximately 5 $\mu$m in radius \cite{Su_2018, Haddrell_2019}.
In acoustic traps, on the other hand, it is difficult to trap droplets below a millimeter due to their low inertia \cite{barrios_dynamics_2008} and high susceptibility to environmental perturbations, such as air currents and thermal fluctuations. 

Optical traps, especially those using counter-propagating beams \cite{Bowman:11}, offer an alternative for studying droplets smaller than 5 \(~\mu\)m in radius. 
These traps have enabled detailed measurements of evaporation rates \cite{Pastel:01}, refractive indices \cite{Rafferty_2018}, Raman scattering \cite{Miles_2012}, and water uptake in organic droplets \cite{Diveky_2021}. 
However, optical traps also face challenges, such as positional instability as droplets evaporate, often leading to ``jumping" between trapping positions \cite{Rafferty_2021}, or being completely lost. 
High-humidity conditions or high salt concentrations in droplets are commonly used to stop the droplets from evaporating and keep them stably trapped.
However, by adding salt the physical properties of the system under investigation are changed.

Despite the importance of understanding droplet evaporation, predicting this behavior in conditions involving radiative heating remains challenging. 
The widely-used $D^2$-Law provides a foundational framework for predicting evaporation rates \cite{dalla_barba_revisiting_2021} by assuming that evaporation is driven by diffusion into the surrounding air, but it does not include radiative heating.
Several numerical and experimental evaporation studies for droplets under laser irradiation have been conducted showing the uneven volumetric absorption of irradiative heat \cite{Dombrovsky_2003} and the evaporation of droplets in air \cite{Pokharel_2022, Zhang_2023}, and on hydrophobic surfaces \cite{Jiao_2016}.
These report only small deviations from the $D^2$-Law with the evaporation rate increasing as they shrink.
However, none of these experiments have investigated the droplets down to full evaporation or under strong irradiative heating.

Here we investigate the evaporation dynamics of water droplets from 10 $\mu$m in radius down to full evaporation. 
We find that droplets early in the evaporation process do not evaporate according to the $D^2$-Law, where $\dot{R}(t)\sim R^{-1}$, but rather as $\dot{R}(t)\sim R$. We attribute these evaporation dynamics to laser-driven evaporation. As evaporation proceeds further and the droplets become smaller, we observe a turnover to diffusion-driven evaporation similar to the classical $D^2$-Law. These findings contribute to a deeper understanding of evaporation processes under strong radiative heating, including atmospheric sciences, internal combustion engines, and aerospace propulsion.


\section{Methods}

\subsection{Counter-propagating trap with IR heating}

\begin{figure}[ht]
\centering
\includegraphics[width=0.9\columnwidth]{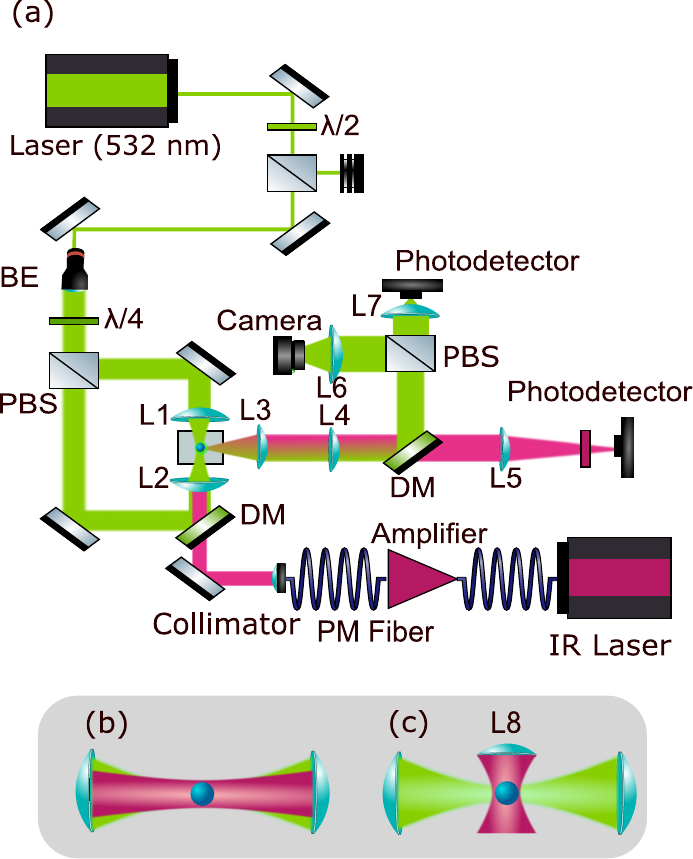}
\caption{Experimental Setup \textbf{(a)} A counter-propagating optical trap levitates water droplets between the lenses L1 and L2. The water droplets are heated with an IR laser beam set up in either configuration (b) or (c). \textbf{(b)} Horizontal configuration where the IR beam enters parallel to the trapping beam and is focused by L1 (f = 50 mm). The long focal distance of L1 leads to a lower power density. \textbf{(c)} Vertical configuration where the IR beam enters perpendicular to the trapping laser. The short focal distance of L8 (f = 25 mm) leads to higher power densities.}
\label{fig:1}
\end{figure}

\begin{figure}[ht]
\centering
\includegraphics[width=0.95\columnwidth]{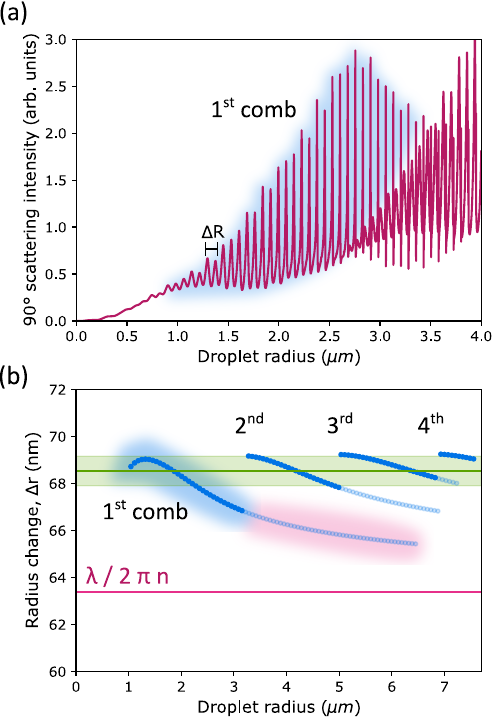}
\caption{We use the radius change of a droplet between resonances in the Fano comb pattern to measure the evaporation rate. 
\textbf{(a)} A simulation of the 90$^{\circ}$ light scattering of a 532 nm plane wave by a water droplet as it evaporates. 
The dominating resonances from the first comb are highlighted in blue.
\textbf{(b)} 
The commonly used approximation that a resonance occurs every time the droplet diameter changes one wavelength inside the material, $\Delta R = \lambda / 2 \pi n$ is plotted in magenta.
The exact radius change between resonances, $\Delta R$, is plotted in blue for the first four combs.
The average radius change, $\Delta R = 68.5 \pm 0.6$ nm, is plotted with a horizontal green line with green shading representing one standard deviation.
The radius changes from the dominating resonances in the first comb are highlighted in blue, while the radius changes from resonances hidden underneath the next comb are highlighted in magenta.
We only use the dominating resonances in this work.}
\label{fig:2}
\end{figure}

We developed a counter-propagating optical trap, shown in \partFig{fig:1}{a}, in which water droplets are trapped and evaporate into the ambient air. 
The laser beam from a  532 nm continuous wave (gem532) was expanded to 4.5 mm in diameter and split into two arms using a polarizing beam splitter. 
Each polarized beam was focused using a pair of 50 mm lenses in a counter-propagating arrangement, with beam waist at a focus of 3.7 ± 0.2 $\mu$m.

This setup was similar to previous optical traps used to study droplet kinetics \cite{Diveky_2021, Rafferty_2018}. 
However, our setup could trap droplets with radii up to 10 $\mu$m and keep them trapped down to full evaporation. 
We achieved this using long focal length lenses instead of microscope objectives in the counter-propagating trap, resulting in larger beam waists. 
On the contrary to other studies where solutes such as salt and high-humidity environment are used to slow or even stop the evaporation \cite{Haddrell_2019,Gregson_2019}, we instead let the droplets evaporate, allowing us to study the evaporation of small droplets without added solutes.

We used an ultrasonic nebulizer (MY-520A) to dispense a cloud of droplets into the chamber.
The droplets fell randomly into the trap and merged up to a maximum droplet size defined by the optical trap.
The continuous dispensing of water vapor resulted in a sustained high relative humidity (RH) inside the chamber between 98 $\pm$ 3\% and an ambient temperature of 297~\text{K}.

To heat the droplets, we used an infrared (IR) laser (TLX1 C-Band Tunable Laser), set to a wavelength of 1550 ± 1 nm, amplified by an erbium-doped fiber amplifier (Thorlabs EDFA100P)  as shown in \partFig{fig:1}(a). 
The fiber output was collimated to a beam diameter of 1.6 mm.

We used two different setups to study the impact of increasing IR irradiation on evaporation dynamics. 
First, a horizontal configuration (\partFig{fig:1}(b)) with a beam waist of 30.3 ± 0.2 $\mu$m, where we scanned the power of the IR laser from 0 to 140 mW in steps of 10 mW.
Second, a vertical configuration (\partFig{fig:1}(c)) with a beam waist of 15.3 ± 0.2 $\mu$m, where we scanned the power from 35 mW (equivalent to 140 mW for the horizontal configuration) to 95 mW in steps of 5 mW.
Importantly, the beam waists were always larger than the droplet radii under study, meaning they always received homogeneous laser heating.
The two setups were operated independently at different times, with the horizontal setup used for lower IR intensities and the vertical setup for higher intensities, allowing us to explore a wide range of IR irradiation levels.

\subsection{Measuring the evaporation rate using the Fano comb structure} \label{subsec: MeasuringWithFano}

Spherical droplets can become resonance cavities for light. 
This happens when the light that evanescently couples into the droplet reflects around the inner water-air interface through total internal reflection.
When the optical path is a multiple of the wavelength inside the material constructive interference occurs and the droplet shines brighter.
These are called Whispering Gallery Modes (WGM) and result in the droplet twinkling as it evaporates.
Fig. \ref{fig:2}(a) shows the 90$^\circ$ scattering intensity of 532 nm light by a water droplet with a refractive index of $n = 1.3355 + 2.6544 \times 10^{-9} i$ \cite{Kedenburg_2012}. 
The scattering shows a Fano comb structure, named so because of the asymmetric Fano profile of the resonances on the far right side of the combs \cite{marmolejo_fano_2023}. 

A common approximation is that resonances occur when the optical path of the light rotating around the droplet is equal to the circumference, $C$.
Hence, resonances occur every time $C = m \lambda$ where  $m$ is an integer.
This approximate change in radius, $\Delta R$, between resonances is plotted in magenta in Fig. \ref{fig:2}(b).
The exact radius change between resonances is plotted in blue in Fig. \ref{fig:2}(b), showing that the approximation slightly underestimates the radius change.
The dark blue line (highlighted in blue for the first comb) marks the radius changes between resonances of a comb that rise above any other comb, i.e. dominant resonances.
The transparent line (highlighted in magenta) marks resonances that fall below those of the next comb. 
The exact turnover will be dependent on the absorbance of the material.
We find that radius change between dominant resonances is approximately constant with an average of \(\Delta R = 68.5 \pm 0.6\) nm, plotted with a green line in  \ref{fig:2}(b).

In our experiments, we record time series of light scattered by evaporating droplets.
To measure the evaporation rate we assume the radius change between resonances to be constant and then measure the period between resonances $\Delta t$. 
Then, the evaporation rate is

\begin{equation}
    \label{eq: drdt_definition}
    \frac{dR}{dt} = \frac{\Delta R}{\Delta t} .
\end{equation}

\subsection{Using the Fano Combs vs. the Far-field Scattering to Measure Evaporation Rates}

Several studies use a recording of the far-field light scattering (also called phase function) of the droplets to measure their size as they evaporate \cite{Su_2018,Diveky_2021,Gregson_2019}.
This has several advantages over using the Fano combs.
To name a few, the sample rate is constant and depends on the frame rate of the camera, it measures the absolute size, and it is a well-established technique.
Both techniques also share disadvantages, including the necessity of knowing a priori the refractive index.
A comparison between these two methods can be found here \cite{Taflin_1998}.

The Mie resonance technique used in this work has several advantages.
Most importantly, for small droplets, the far-field scattering can become so large that very few if any fringes are visible.
This depends on the numerical aperture (NA) of the lens collecting the scattering of the droplets.
However, for applications where the light cannot be collected at short range, the Mie resonances technique provides access to measurements of droplets below radii of 5 $\mu m$. 

Secondly, it only requires a photodiode instead of a camera. 
This, on top of reducing the cost of the experiment, results in a measurement made up of a one-dimensional time series instead of a video, increasing the accessible sampling rate.
Using the resonances we achieved a maximum sampling rate of 1 400 sps, which, although reachable with a high-speed camera, is two orders of magnitude faster than a normal 30 fps camera and has a much lower memory cost.



\section{Theory}

\subsection{$D^2$-Law Evaporation}

The classical $D^2$-Law of droplet evaporation \cite{d2doi:https://doi.org/10.1002/9781118916643.ch4, LAW1982171} describes how the size of a spherical droplet decreases over time, assuming spherical symmetry for both the droplet and surrounding gas. For this law to be valid, several assumptions must hold: the evaporation process is limited by diffusion of water molecules into the ambient air, the droplet maintains a constant temperature, density, and surface tension, and remains spherical throughout the process. Additionally, the surrounding gas must be stagnant, with no external forces acting on the droplet, and mass transfer should be unaffected by chemical reactions, condensation, or coalescence. The model further assumes that the ambient vapor pressure is much lower than the droplet’s saturation vapor pressure, the droplet’s surface remains stationary relative to vapor diffusion, and that radiative heat transfer is negligible.

A generalized form of the $D^2$-Law reads \cite{dalla_barba_revisiting_2021}
\begin{equation}
    \label{eq:d2}
    D^2(t) = D_0^2 - K t,
\end{equation}
\noindent
where \(D(t)\) is the droplet diameter at time \(t\), \(D_0\) is the initial diameter, and \(K\) is the evaporation rate constant. Changing the diameter to radius, R, and derivating \Eq{eq:d2} as a function of time we obtain
\begin{equation}
    \label{eq:drdt_ambient}
    \frac{dR}{dt} = - \frac{K}{8R(t)} = -\frac{K}{4 \sqrt{D_0^2 - K t}}.
\end{equation}
Eq. (\ref{eq:drdt_ambient}) highlights the dependence for the diffusion-limited evaporation rate as a function of time and radius: power laws proportional to $t^{-1/2}$ and to $R^{-1}$.

The evaporation constant $K$ depends on characteristic properties of the droplet and the ambient air \cite{dalla_barba_revisiting_2021} and can be generally defined as
\begin{equation}
    \label{eq: K}
    K = \frac{\rho_a}{\rho_l} \cdot \frac{\text{Sh}}{\text{Sc}_a} \cdot H_m \cdot v_a,
\end{equation}
where the parameter
\begin{equation}
    H_m = \ln \left( \frac{1 - Y_{v,m}}{1 - Y_{v,d}} \right),
\end{equation}
\noindent
describes the mass transfer rate. 
We note that the parameter $H_m$ is expected to change with the pressure and temperature of the system \cite{dalla_barba_revisiting_2021}, which we assume to be approximately constant during the diffusion-driven regime in our experiments. 
In these equations, \(\rho_a\) and \(\rho_l\) are the densities of the ambient gas and the liquid in the droplet respectively.
\(Sh\) is the Sherwood number which compares the convective mass transfer to the mass transfer by diffusion and \(Sc_a\) is the Schmidt number of air for water vapor which compares the viscous diffusion rate to the mass diffusion rate.
\(T_a\) is the ambient temperature, \(p_a\) is the ambient vapor pressure, \(RH_a\) is the relative humidity of ambient, and \(v_a\) is the dynamic viscosity of air.
\(Y_{v,m}\) is the vapor mass fraction at the water-air interface, and \(Y_{v,d}\) is the mass fraction of a saturated gas–vapor mixture at a given droplet temperature.

\subsection{Irradiation-Driven Evaporation}

We will now account for the irradiative heating on the droplets.
We will use several strong assumptions, which, although not valid for the whole evaporation process, illustrate the effect of strong irradiation and help interpret the data shown in the following section.

The assumptions are the following: 
(1) The droplet under strong irradiative heating reaches a steady state with constant temperature where all the heat from the lasers is lost through evaporation, as shown in  \cite{Pokharel_2022} where the temperature reaches an approximately constant level.
(2) The heat loss through evaporation is so dominant that the convective and radiative heat losses become negligible.
(3) The droplet maintains a spherical size and a constant environment.
With these approximations, the heat balance equation becomes
\begin{equation}
    \label{eq:balance_gen}
    \dot{Q}_{\text{abs,trap}} + \dot{Q}_{\text{abs,IR}} = \dot{Q}_{\text{evap}} ,
\end{equation}
where \(\dot{Q}_{\text{abs,trap}}\) and \(\dot{Q}_{\text{abs,IR}}\) are the heat absorption rates from the trapping and IR lasers. \(\dot{Q}_{\text{evap}}\) is the heat loss due to evaporation which depends on the latent heat of vaporization $L$ and the mass evaporation rate \(\dot{m}\) as
\begin{equation}
    \dot{Q}_{\text{evap}} = \dot{m} L .
\end{equation}
\noindent

The mass evaporation rate \(\dot{m}\) can be related to the radius change as
\begin{equation}
    \dot{m} = -\rho_l \frac{dV}{dt} = -\rho_l \cdot 4 \pi R^2 \frac{dR}{dt}.
\end{equation}
Substituting this into Eq. (\ref{eq:balance_gen}) and solving for the evaporation rate, we obtain
\begin{equation}
    \label{eq:dr_dt_2}
    \frac{dR}{dt} = -\frac{\dot{Q}_{\text{abs,trap}} + \dot{Q}_{\text{abs,IR}}}{4 \pi \rho_l L R^2}.
\end{equation}

For both the optical trap and the IR laser, the heat absorption rate $\dot{Q}_{\text{abs}}$ of the droplet is given by
\begin{equation}
\label{eq:Q_abs_1}
    \dot{Q}_{\text{abs}} = q_{\text{abs}} \cdot I_{\text{laser}} \cdot A_{\text{droplet}},
\end{equation}
where \(q_{\text{abs}}\) is the Mie absorption efficiency, \(I_{\text{laser}}\) is the irradiation of the laser and \(A_{\text{droplet}}\) is the droplet cross sectional area. Using $I_{\text{laser}} = P_{\text{laser}}/A_{\text{beam}}$, where \( P_{\text{laser}} \) is the laser power and \( A_{\text{beam}} \) is the cross-sectional area at the waist of the beam and $A_{\text{droplet}}=4\pi R^2$, \Eq{eq:Q_abs_1} can be further expressed as
\begin{equation}
\label{eq:Q_abs}
    \dot{Q}_{\text{abs}} = q_{\text{abs}} \cdot \frac{P_{\text{laser}}}{A_{\text{beam}}} \cdot \pi R^2.
\end{equation}

Here, \( q_{\text{abs}} \) represents the absorption efficiency of light by the droplet, as derived from Mie theory. In the geometrical optics approximation, this efficiency is proportional to the droplet radius, as detailed in \cite{bohren2008absorption}.
\(q_{\text{abs}} = \gamma R\), with $\gamma = (4 / 3) (\alpha / (n_{wter}/n_{air})) [ (n_{wter}/n_{air})^3 - \left( (n_{wter}/n_{air})^2 - 1 \right)^{3/2} ]$ a constant dependent on the refractive indices, and the absorption coefficient of the droplet, $\alpha$ \cite{bohren2008absorption}.

Then, by substituting \Eq{eq:Q_abs} into \Eq{eq:dr_dt_2} we obtain
\begin{equation}
    \label{eq:drdt_laser}
    \frac{dR}{dt} = -\frac{R (P_{trap} \gamma_{trap} + P_{IR} \gamma_{IR})}{4 A_{\text{beam}} \rho_l L}.
\end{equation}
\Eq{eq:drdt_laser} highlights that in the laser-driven regime, we expect a power law relationship between the evaporation rate and the radius proportional to $R$.

\section{Results}

\subsection{Diffusion-driven evaporation}

\begin{figure*}[ht]
\centering
\includegraphics[width=\textwidth]{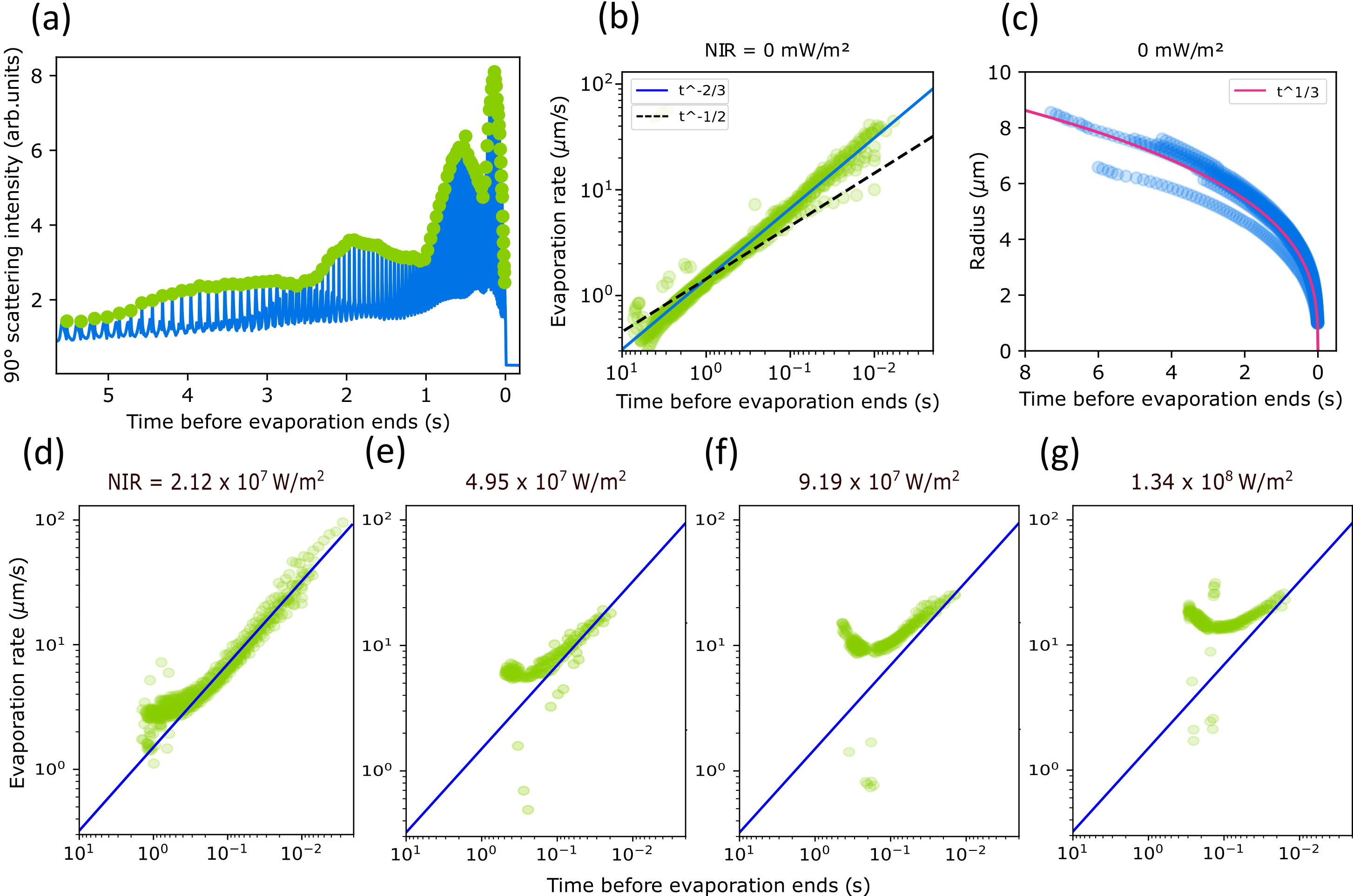}
\caption{The evaporation rate follows a $\dot R \sim t^{-2/3}$ power law in the absence of IR heating. In the presence of IR heating, the evaporation rate tends towards this power law at the end of the evaporation. 
\textbf{(a)} Fano combs of an evaporating water droplet in a laser trap without IR irradiation. The green dots mark the resonances used to calculate the evaporation rate following the method described in Sec. \ref{subsec: MeasuringWithFano}. 
\textbf{(b)} The evaporation rates of ten separate water droplets in a laser trap (green dots) follow a $\dot R \sim t^{-2/3}$ power law plotted with a blue line. The classical $D^2$-Law $ \dot R \sim t^{-1/2}$ is plotted with a dashed black line for comparison. 
\textbf{(c)} Integrating the evaporation rates from (b) we obtain the droplet radius during the evaporation. It follows $R \sim t^{1/3}$ behavior instead of $R \sim t^{1/2}$ given by $D^2$-Law. 
\textbf{(d)-(g)} Evaporation rates at increasing IR irradiation levels, each showing several separate droplets overlapped. Here solid blue line shows the same $\dot R \sim t^{-2/3}$ power law as in (a).
}
\label{fig:3}
\end{figure*}

Fig. \ref{fig:3}(a) shows an example of the twinkling of a water droplet as it evaporates without any IR heating. 
This light scattering is perpendicular to the illuminating laser and forms a Fano comb pattern.
The dominating resonances are marked with green dots.

We then use the period between resonances to measure the evaporation rate shown in Fig. \ref{fig:3}(b). 
Here we plotted the evaporation rates of 10 separate droplets on a log-log scale. 
Note also that the x-axis is inverted showing instead the ``time before evaporation ends'', $t'$.
The evaporation rate for these droplets increases polynomially as the droplet shrinks.
The blue line shows a best-fit curve for the data from all 9 droplets together using the equation 
\begin{equation}
    \label{eq: D-2/3}
    \frac{dR}{dt} =  B (t')^{-2/3} ,
\end{equation}
where B is a fitting constant. Eq. (\ref{eq: D-2/3}) can be written as 
\begin{equation}
    \label{eq: D-2/3-equivalent}
    \frac{dR}{dt} = B (t_0 - t)^{-2/3} ,
\end{equation}
using instead an initial time, $t_0$, and time, t.
The black dashed line illustrates the trend of the $D^2$-Law for comparison.
The fact that these power laws appear increasing even though the exponents are negative is a consequence of the inversion of the x-axis to plot the time before evaporation ends.
Fig. \ref{fig:3}(c) shows the same data but plotting the radius instead, which follows a $R^{1/3}$ power law.

It is surprising that the evaporation dynamics without additional IR irradiation follows a $t^{-2/3}$ power law, which suggests that in this experiment the droplet is being heated by the optical trap. Even though the absorption of the 532 nm trapping laser is small (for comparison, the absorption of 1550 nm by water is 5 orders of magnitude larger), it may produce non-uniform heating since the droplet is larger than the laser beam waist, which could potentially result in the observed deviations from the $D^2$-Law.
Another reason could be the small drift of the position of the droplet inside the laser trap towards zones of higher laser power, which also causes it to shine brighter as it shrinks.

\subsection{Irradiation-driven evaporation}

Figs.\ref{fig:3} (d - g), show again an overlap of the evaporation rates of ten different droplets, where now each has an increasing level of IR irradiation. For low irradiation levels, the evaporation follows the same $t^{-2/3}$ power law as in \ref{fig:3}(b), as indicated by the solid blue line.
However, on the left side of the figures showing the beginning of the evaporation, the evaporation rate becomes increasingly higher.
Furthermore, for the highest irradiation levels, Figs.\ref{fig:3} (f) and (g), a bend appears where the evaporation rate first drops and then increases again.

This same behavior is shown again in Fig. \ref{fig:4} as a function of radius instead of time.
The blue dashed line illustrates the $D^2$-Law as a function of radius, i.e. $\sim R^{-1}$ from Eq. (\ref{eq:drdt_ambient}), as a reference of diffusion-driven evaporation.
The magenta dashed line represents the irradiation-driven evaporation, i.e. $\sim R$ from Eq. (\ref{eq:drdt_laser}).
This result shows that for smaller droplets diffusion-driven evaporation is dominant while for larger droplets irradiation-driven evaporation is dominant.

The constants related to each power law are $\beta K$ and $ -(P_{trap} \gamma_{trap} + P_{IR} \gamma_{IR}) / 4 A_{\text{beam}} \rho_l L$, where $\beta$ is a fitting constant that accounts for droplet heating in the diffusion-driven regime. 
These were calculated using Eq. (\ref{eq: K}) and the values shown in the Supplemental Material \footnote{See Supplemental Material at [URL will be inserted by publisher] including reference \cite{Wang_2021}}.

\subsection{Turnover from irradiation-driven to diffusion-driven evaporation}

The turnover we observe between the two regimes starts at radii of approximately \(3 \, \mu\text{m}\) at lower irradiation levels and decreases to approximately \(2.5 \, \mu\text{m}\) in radius at higher irradiation levels. 
The minimum evaporation rate at the turnover increases progressively up to $\sim 10 \mu m / s $, and the bend becomes more significant as the irradiation increases. 

\begin{figure*}[ht]
\centering
\includegraphics[width=\textwidth]{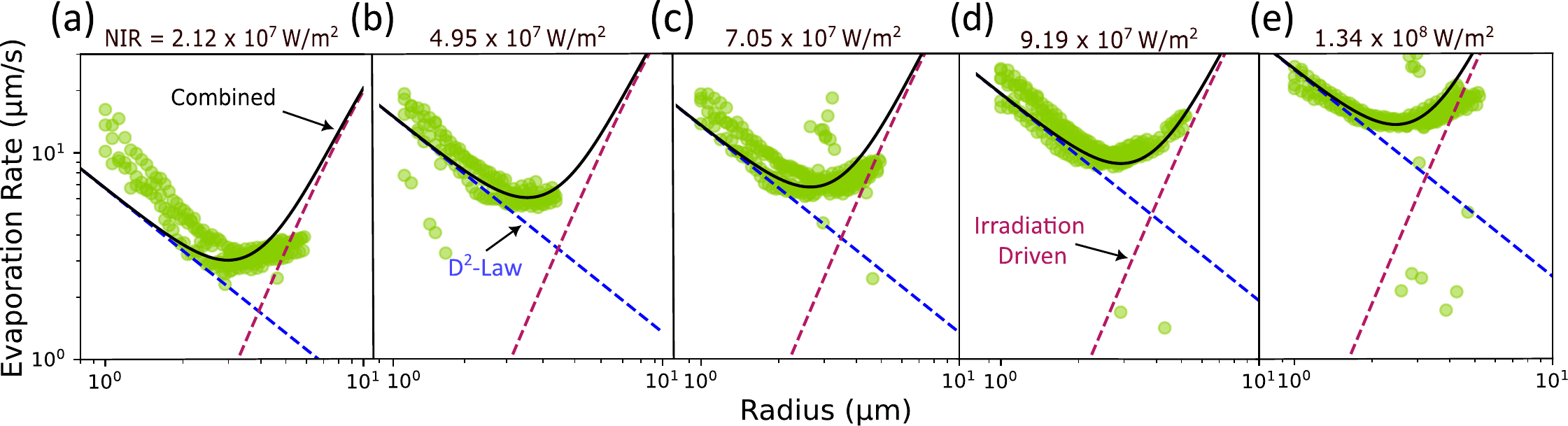}
\caption{Turnover from irradiation-driven (magenta dashed line) to diffusion-driven (blue dashed line) evaporation. 
\textbf{(a)-(e)} The evaporation rates of 10 droplets at increasingly stronger IR irradiation. The black line shows a fit using a linear combination between the two regimes described in \Eq{eq:drdt_ambient} and \Eq{eq:drdt_laser}.}
\label{fig:4}
\end{figure*}

\Fig{fig:4} shows our model of this turnover where we use a linear combination of the terms in \Eq{eq:drdt_ambient} and \Eq{eq:drdt_laser}. 
We observe that the coefficient for the $D^2$-Law term, $\beta$, in this combination must increase with higher IR irradiation to match the experimental data.
This means that irradiative heating also affects the diffusive evaporation regime, most likely by increasing the maximum temperature of the droplet. 

This aligns with the revised $D^2$-Law proposed in \cite{dalla_barba_revisiting_2021}, which introduces a droplet asymptotic temperature \(T_{ds}\) in the expression of the evaporation constant \(K\).
Although, in their model the droplet asymptotic temperature \(T_{ds}\) cannot be greater than the ambient temperature, the external heating from the IR laser makes it possible for the \(T_{ds}\) to be greater than the ambient temperature \(T_a\). 
We suspect that this temperature \(T_{ds}\) increases as the laser irradiation intensifies, which in turn increases the value of the evaporation constant \(K\).

\subsection{Heating during a WGM resonance}

\begin{figure}[ht!]
\centering
\includegraphics[width=0.9\columnwidth]{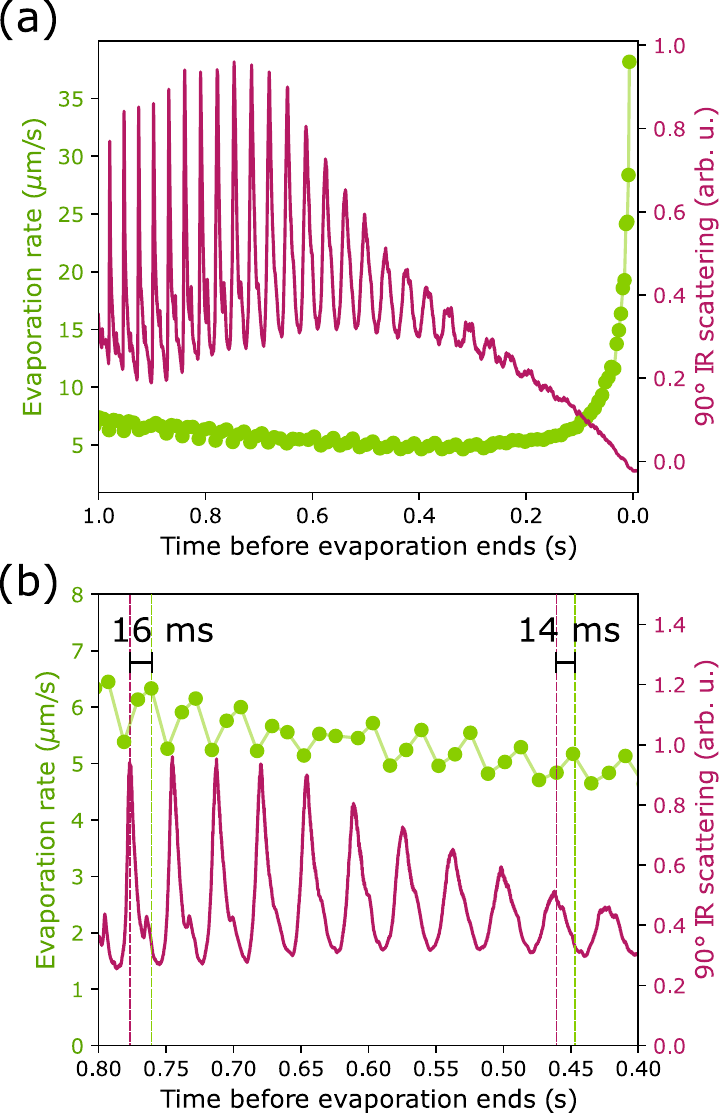}
\caption{Fast increase in the evaporation rates of a droplet following a Mie resonance. \textbf{(a)} Mie combs in the IR scattering of the droplet (magenta, right y-axis) and the evaporation rate  (green dots, left y-axis) \textbf{(b)} Zoomed-in plot of the evaporation rate showing the increases caused by the increased absorption because of the Mie resonances.}
\label{fig:5}
\end{figure}

We also measured the IR scattering, shown in \Fig{fig:5}(a), simultaneously to the measurement of the 532 nm laser used to calculate the evaporation rate.
The zoom-in to a 400 ms period in \Fig{fig:5}(b) shows that the evaporation rate oscillates from fast to slow in correlation with the IR resonances.
Since we use the resonances to measure the evaporation rate and the IR wavelength is about three times longer than the trapping wavelength, we have three evaporation rate measurements per IR resonance.

The increased heat absorbance during Mie resonances is well-known \cite{bohren2008absorption}.
However, we can use this fact showcased in \Fig{fig:5}(b) to illustrate how the evaporation rate can adapt fast to changes in irradiation.
We observe increases of 10 - 15\% in evaporation rate with a delay of 14 - 16 ms between the Mie resonance and the peak evaporation speed.

\section{Discussion and Conclusion}

Our work showcases the strong effects of irradiation on the evaporation rate of micrometer-sized water droplets.
We find that early in the evaporation process, when the droplets are larger than 3 $\mu$m in radius, strong irradiative heating drives the evaporation dynamics, which become different from the classical $D^2$-Law.  In this regime the evaporation rate decreases as $\dot{R}(t) \sim R$. 
As evaporation proceeds, we observe a turnover in the evaporation dynamics after which the evaporation returns to a diffusion-driven behavior where $\dot{R}(t) \sim R^{-1}$. 

We can explain the turnover between irradiation-driven evaporation and diffusion-driven evaporation by considering the dimensionality of the system.
Diffusion-driven evaporation is a surface effect proportional to the area of the droplets, while, since the droplets are transparent, the heating from the irradiation is a volume effect.
For larger droplets the irradiative heating is large, meaning that it heats up the droplet and increases the evaporation rate until the evaporation can remove heat comparable to the irradiative heating it receives.
As the droplet gets smaller, the irradiative heating decreases cubically while the surface area decreases quadratically. 
Eventually, the irradiative heating becomes negligible and the evaporation becomes again diffusion-driven.



In conclusion, in this paper we showed and explained the effects of strong irradiation on evaporating droplets with radii below 10 $\mu m$. 
The Mie resonance technique we used to calculate the evaporation rates allowed us to measure evaporation rates accurately and with a high sampling rate.
With our optical trap, we were able to study spherical, levitating drops with no lower limit on droplet size, which are not in thermal contact with other materials. This setup allowed us to study the effects of laser irradiation on the evaporation rate in a controlled manner.
We explained our experimental results through simple thermodynamic theory, proving the validity of our approach. 
Our findings highlight how irradiative heating can affect the evaporation dynamics of small droplets, which is important for understanding evaporation dynamics in many applications in nature and industry, for instance in internal combustion engines like gas-turbines where the irradiation from the flame heats up fuel droplets.

\section{Author Contributions}
J.T.M. conceptualized the experiment. J.R.S. and J.T.M. built the experimental setup, performed the experiments, and analyzed the data. J.R.S., J.T.M., and M.H. wrote the first draft of the manuscript. J.R.S. and M.H. contributed to the interpretation of the data and the theoretical background. D.H., D.D., and J.T.M. supervised the project. All authors contributed to the final version of the manuscript.

\section{Acknowledgments}
Financial support from the Swedish Research Council (2019-02376 and 2020-03505) is acknowledged.
J.T.M. acknowledges support from The Märta and Eric Holmberg Endowment and Stiftelsen Lars Hiertas Minne. M.H. was funded by the University of Edinburgh. J.R.S. acknowledges support from Erasmus+ programme of the European Union. We thank Ingemar Magnusson, Simon Titmuss, and Wilson Poon for discussions regarding the interpretation of our results.



\bibliographystyle{unsrt}
\bibliography{References}

\end{document}


\title{Supplemental Material \\
Irradiation-driven Evaporation of Micro Droplets in an Optical Trap}

\author[1]{Jugal Rakesh Shah}
\author[2]{Max Huisman}
\author[1]{Devendra Deshmukh}
\author[3]{Dag Hanstorp}
\author[3]{Javier Tello Marmolejo}
\date{}

\affil[1]{Department of Mechanical Engineering, Indian Institute of Technology, Indore 452017, MP, India\\}
\affil[2]{SUPA and School of Physics and Astronomy, The University of Edinburgh, Peter Guthrie Tait Road, Edinburgh EH9 3FD, United Kingdom\\}
\affil[3]{Department of Physics, University of Gothenburg, SE-412 96 Gothenburg, Sweden\\}

\maketitle

\section*{Constants and Parameters}
The following constants and parameters were used to calculate the evaporation rate constant, $K$, in Eq. 4 and the evaporation rates below.

\begin{multicols}{2}
\begin{itemize}
    \item \textbf{Refractive Indices and Absorption Coefficients} \cite{Kedenburg_2012}
    \begin{itemize}
        \item Water (532 nm): $n = 1.3355 + 2.6544 \times 10^{-9}i$, $\alpha = 0.000345 \, \text{cm}^{-1}$
        \item Water (1550 nm): $n = 1.3189 + 9.8625 \times 10^{-5}i$, $\alpha = 7.9959 \, \text{cm}^{-1}$
    \end{itemize}
    \item \textbf{Densities:}
    \begin{itemize}
        \item Air: $\rho_a = 1.19 \, \text{kg/m}^3$
        \item Water: $\rho_l = 997 \, \text{kg/m}^3$
    \end{itemize}
    \item \textbf{Environmental Conditions:}
    \begin{itemize}
        \item Ambient Temperature: $T_a = 297 \, \text{K}$
        \item Relative Humidity: $\text{RH}_a = 98\%$
    \end{itemize}
    \vspace{2cm}
    \item \textbf{Other Properties:}
    \begin{itemize}
        \item Latent Heat of Vaporization of water: \\ $L = 2257.06 \, \text{kJ/kg}$
        \item Dynamic Viscosity of air: \\$v_a = 1.562 \times 10^{-5} \, \text{kg/ms}$
        \item Kinematic viscosity of air (20°C): $\bar v = $ 0.0000155 $m^2/s$
        \item Mass diffusivity of water vapor in air (20°C): $D_{g,v} = $ 0.0000242 $m^2/s$
        \item Schmidt Number: $\text{Sc}_a = \bar v / D_{g,v} = 0.64279$
        \item Sherwood Number: $\text{Sh} = 2$ \\
        Using the Frössling correlation, $Sh_0 = 2 + 0.552Re_d^{1/2}Sc^{1/3}$ \cite{Wang_2021} and assuming $Re_d \ll 1$, which is reasonable for a micron-sized droplet evaporating into stagnant air.
        
    \end{itemize}
\end{itemize}
\end{multicols}

\vspace{0.3cm}

\section*{Equations}
\noindent The dashed lines showing the evaporation rate in Fig. 4 were derived using the following equations:

\noindent \textbf{Diffusion-driven term:}
\begin{equation*}
    \frac{dR}{dt} = - \frac{K}{8R(t)} = -\frac{K}{4 \sqrt{D_0^2 - K t}}.
\end{equation*}

\noindent \textbf{Laser-driven term:}
\begin{equation*}
    \frac{dR}{dt} = -\frac{R (P_{trap} \gamma_{trap} + P_{IR} \gamma_{IR})}{4 A_{\text{beam}} \rho_l L}.
\end{equation*}

\noindent \textbf{Combined term using Linear Combination:}
\begin{equation*}
    \frac{dR}{dt} = - \beta \frac{K}{8R(t)} -\frac{R (P_{trap} \gamma_{trap} + P_{IR} \gamma_{IR})}{4 A_{\text{beam}} \rho_l L},
\end{equation*}

where $\beta$ is a fitting constant that accounts for minor droplet heating in the diffusion-driven evaporation regime.

\vspace{0.2cm}

\section*{Table of Constants}

\noindent The fitting constants for the diffusion term in the linear combination were

\begin{table}[h!]
    \centering
    \small 
    \begin{tabular}{c c c}
        \toprule
        \textbf{Irradiation (W/m$^2$)} & \textbf{$\beta$} & \textbf{Resulting evaporation constant} $\beta K$ \\
        \midrule
$2.12 \times 10^{7}$ & 3.5 & $5.375173 \times 10^{-11}$ \\
$4.95 \times 10^{7}$ & 7 & $1.075034 \times 10^{-10}$ \\
$7.05 \times 10^{7}$ & 8.5 & $1.305399 \times 10^{-10}$ \\
$9.19 \times 10^{7}$ & 10 & $1.535763 \times 10^{-10}$ \\
$1.34 \times 10^{8}$ & 13 & $1.996492 \times 10^{-10}$ \\
\bottomrule
\end{tabular}
    \caption{Variation of constant $C$ with irradiation values.}
    \label{tab:parameters}
\end{table}

\vspace{0.2cm}
\noindent These constants and equations form the basis of the theoretical framework used to analyze evaporation dynamics.

\bibliographystyle{unsrt}
\bibliography{References}